# Power Converters for Accelerators

*R. Visintini*
Elettra Sincrotrone Trieste, Trieste, Italy

**Abstract**
Particle accelerators use a great variety of power converters for energizing their sub-systems; while the total number of power converters usually depends on the size of the accelerator or combination of accelerators (including the experimental setup), the characteristics of power converters depend on their loads and on the particle physics requirements: this paper aims to provide an overview of the magnet power converters in use in several facilities worldwide.

## 1    Introduction

According to some recent statistics (2011) [1], there are about 30 000 particle accelerators operating in the world, mostly used for industrial (20 000) and medical (10 000) purposes. Scientific research applications are a small fraction of the total [2], as can be seen in Fig. 1 (data extracted from Ref. [3]).

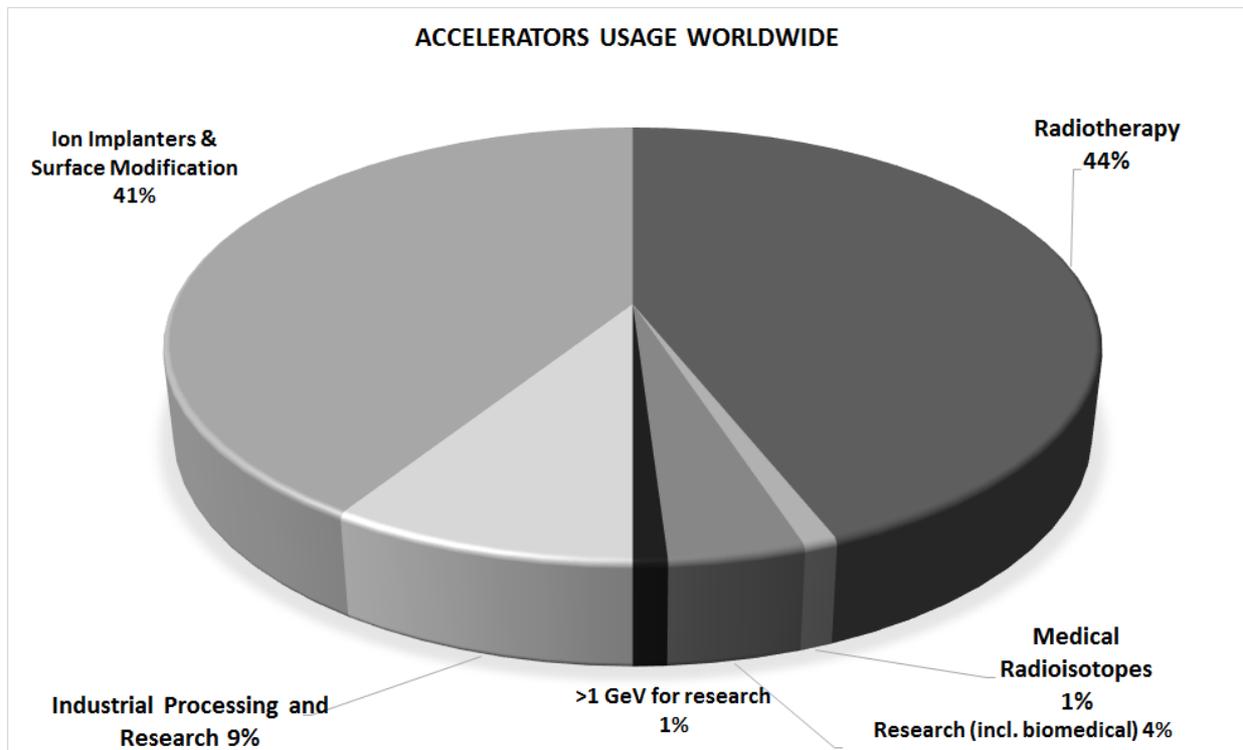

**Fig. 1:** Accelerators usage worldwide (2011)

Compared to the other applications, particle accelerators for scientific research quite often require 'close-to-the-edge' technologies and custom-designed equipment. Such devices (or solutions derived/inspired from them) can also find uses in industrial or medical applications.

## 1.1 Aim of this paper

The subsystems of accelerators, either a relatively small piece of equipment or a large structure, require a variety of power converters of different type, size, and performance. To somewhat limit the vastness of the field, only the accelerators related to scientific applications will be considered here.

During the specialized CAS course on power converters in 2004, H.-J. Eckoldt presented a lecture entitled 'Different power supplies for different machines'. In his paper [4], focused on magnet power converters, he provided a good overview of technologies, topologies, and magnet connections to power converters in special applications, with many examples of solutions adopted by several facilities worldwide.

In this paper, I will try to integrate Eckoldt's work, by presenting a comparison of the requirements of power converters (PC) for magnets, according to the different applications for particle accelerators (PA).

## 2 Particle accelerators

Several sources report the history of particle accelerators (see, for example, Ref. [5]). The first artificial source of accelerated 'high' energy particles (or, better, particles with energies higher than those obtainable from natural sources) is the accelerator constructed by J.D. Cockcroft and E.T.S. Walton in 1930 at the Cavendish Laboratory in Cambridge, England. It has to be noted that the researcher, Walton himself, as shown in Fig. 2, was sitting extremely close to the interaction point between particles and target (radiation issues were still to be investigated…).

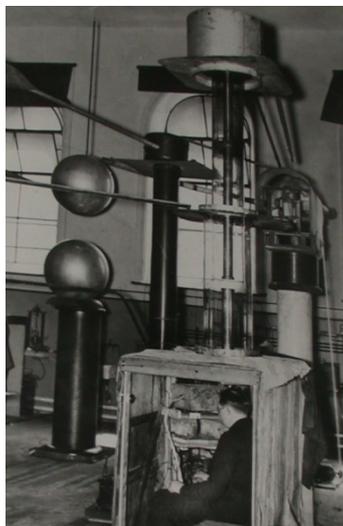

**Fig. 2:** Photograph of the first Cockroft-Walton machine, 1932 (Credit: Cavendish Laboratory, University of Cambridge).

The Cockroft-Walton machine was an electrostatic device, based on a voltage multiplier structure. Almost in parallel (1931), accelerators based on RF oscillators appeared: the linear accelerator (D. Sloan and E.O. Lawrence, starting from the work of R. Wideröe and G. Ising) and the first circular machine, the cyclotron, invented by E.O. Lawrence; see Figs. 3 and 4.

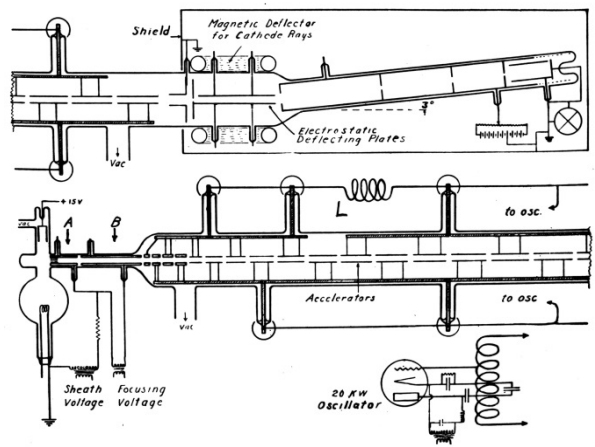

**Fig. 3:** Diagram of Sloan-Lawrence linear accelerator, (Credit: Lawrence Berkeley National Laboratory)

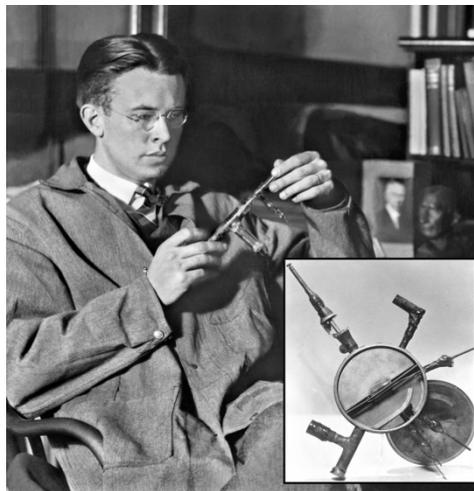

**Fig. 4**: E.O. Lawrence and his 5-inch cyclotron, the first successful cyclotron built by Lawrence and his graduate student M. Stanley Livingston, (Credit: Lawrence Berkeley National Laboratory).

These machines, along with the betatron and synchrotron, are the 'common ancestors' of the modern accelerators; and power converters played a key role since the very beginning.

## 2.1    Particle accelerator classification

One could organize the PA types in various ways. For the purposes of this paper, I consider the following structure:

–    'linear' or 'open', like LINACs or FELs;

–    'circular' or 'closed', like cyclotrons or synchrotrons.

The specific application is also an important criterion that is relevant for the required characteristics of the power converters:

–    high energy physics colliders (HEP-C);

–    ion sources/cancer therapy (IS/CT);

–    neutron sources (NS);

–    light sources (LS).

Independently from the structure—open or closed—and from the specific field of application there are some common actions between the different types of PA:

–   production of particles;

–   acceleration (increasing the energy) of particles;

–   'handling' of particles;

–   measure the energy of particles.

## 2.2    Particle accelerators and power converters

The actions mentioned above are performed using specialized and dedicated equipment or complete subsystems that use power converters whose output could either be a current or a voltage; AC, DC or pulsed; high current, high voltage or both; low power—watts or kW, or high power—hundreds of kW or MW.

As anticipated in the Introduction, for several reasons magnet power converters are the object of this paper, beyond the obvious requirements of 'limiting' the field:

–   in my opinion, the world of magnet power converters is fascinating[1];

–   magnets are everywhere in a PA (magnetic lenses on guns, focusing coils on klystrons, solenoids on accelerating structures, magnets and coils, compensation or correcting coils on insertion devices, spectrometers, etc.);

–   magnets can be normally conducting (warm) or superconducting (cold);

–   magnets can be DC operated, AC operated (often with 'exotic' waveforms) or 'pulsed'.

### 2.2.1    The role of magnet power converters

Particle physicists (or 'machine physicists') study and define the characteristics of the particle beams according to the application (e.g. collider or light source). During and especially after the 'acceleration phases' of the particles, magnetic fields are normally used to 'shape' and drive the particle beams. Consequently, the characteristics of the magnetic fields have a great influence on the 'good quality' of the particle beams.

When electromagnets are used, there are two major issues to consider: the electromechanical design (materials, profile of the poles, shape of the coils, etc.), and the excitation current (ripple or harmonic content, stability, reproducibility, etc.). The power converters energizing the magnets play a key role in matching the required performance of the accelerator.

### 2.2.2    Physics, magnetics, power, control, plant: a system

A particle accelerator is a complex system: its performance also depends upon environmental constraints, like the stability of the ambient temperature where it is installed as well as the temperature of the rooms/galleries where the equipment—power converters, diagnostic electronics, etc.—are located.

Energy considerations—in particular for large facilities—have become strategic issues (see, for example, Refs. [6] and [7]). Magnets and the associated power converter designs are strictly interconnected activities, along with the specifications of the cables connecting them. Adopting a proper cross-section for the cables reduces the voltage drop on the cables, the output power required from the power converters and—from an operational point of view—allows for the reduction of costs.

---

[1]Once, when asked about my job at Elettra, I answered: "I'm providing particle physicists a tool to manipulate with micrometric precision, without any contact, almost immaterial and invisible particles travelling close to the speed of light".

Any watts dissipated in the magnet, power converter and cables are paid for twice: once from the mains to provide them and once from the cooling plant to remove them. Installation issues are also to be considered. Figure 5 summarizes some of the relevant interconnections among the different 'players'.

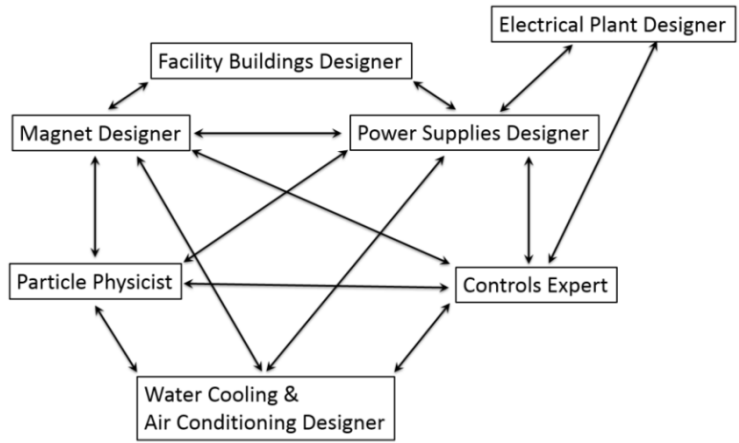

**Fig. 5**: Magnet, power converter, control, plant: a system

### 2.2.3    Some definitions

Before starting a detailed comparison of the characteristics required by the different applications among PAs, I want to recall some useful definitions.

#### 2.2.3.1    Current stability

Current stability is a measure of long-term drift (a percentage of full-scale), over several hours at fixed line, load and temperature, after a warm-up period. Figure 6 shows a real-life example [8] of a stability test over more than 7 hours on a magnet power converter (750 A, 25 V) currently in use with the FERMI FEL source (see, for example, Refs. [9, 10]) at the Elettra Laboratory.

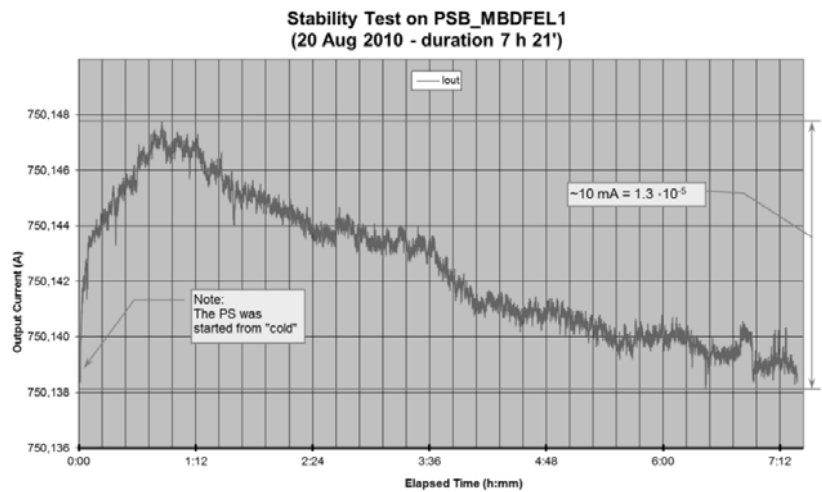

**Fig. 6**: Example of current stability test

#### 2.2.3.2    Current ripple

This is noise on the output current specified as a percentage of the full scale. The frequency spectrum depends on the technology adopted and frequency of commutation of the switches.

This is the smallest possible steps for adjustment of the current set point or the current read-back, specified as a percentage of the full scale or number of bits.

### 2.2.3.4   *Reproducibility*

Reproducibility of the actual output current, for the same current set point (at different times) of a desired output value under constant conditions is specified as a percentage of the full scale.

### 2.2.3.5   *Accuracy (set and read-back)*

Accuracy is a measure of how close the actual output current is to the current set point or to the current read-back, specified as a percentage of the full scale.

## 3      Particle accelerators and magnet power converters

In the following sections I will describe many examples of magnet power converters in use in facilities worldwide, according to my classification as described in Section 2.1. Most of the information is derived from private communications and openly available sources, such as proceedings of particle accelerator conferences (see Ref. [11] for a complete list) or the web sites of the facilities. Images shown in the following sections have been taken from the facilities' web sites or free/open internet image banks through the courtesy of each facility. I have reported the photo credits, when available and required, in the captions.

The particle accelerators for scientific research applications are a small fraction of the total, nevertheless they are sufficiently numerous and the variety of power converter in use is so big to force me to limit the number of examples. I therefore apologize if when reading this paper you do not find your facility.

The examples refer to operational facilities or facilities under construction (at the time of writing) and the dates by their names indicate the start of commissioning/operation. I have tried to structure tables as uniformly as possible but there are differences between them due to the variety of formats adopted in the various documents and sources in describing the power converters and the available data. When not explicitly indicated, I have gathered the data reported via private communications.

### 3.1   High energy physics colliders (HEP-C)

The facilities in this field of applications are characterized by very high particle energies, in the Tera electronVolt (TeV) region) and very large dimensions (kilometres in circumference). There is a large use of superconducting magnets (and conventional ones, too); a very large number of magnet power converters of all types: high current, high voltage or both; low power—watts or kW—or high power—hundreds of kW or MW.

### 3.1.1     *LHC (CERN, Switzerland, 2008 to 2009)*

The LHC is the largest accelerator—actually a complex of accelerators, see Fig. 7—featuring several 'world records':

–   highest particle energy, 7 TeV (3.5 TeV + 3.5 TeV);

–   largest structure (27 km in circumference);

–   largest cryogenic system in the world and one of the coldest places on Earth, main magnets operating at a temperature of 1.9 K (−271.3°C) [12];

- largest number of superconducting magnets (~9600);
- and more…

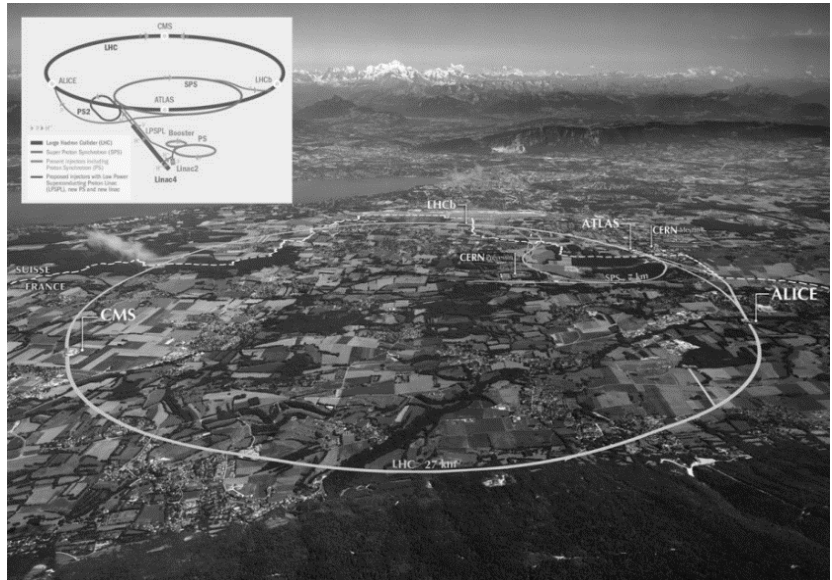

**Fig. 7**: The LHC (credit: CERN)

There are more than 1700 magnet power converters, either unipolar or bipolar (two- or four-quadrant), with currents as high as 13 kA. The technologies adopted are both thyristor rectifiers (silicon-controlled rectifiers - SCR) and switch mode (SM) with insulated-gate bipolar transistors (IGBTs) or metal oxide semiconductor field-effect transistors (MOSFETs). For large output current power converters, smaller units are connected in parallel. The power converters (PCs) are custom-made one, often collaborating with industries in developing the PCs. Since almost all PCs are located underground and are not easily accessible for maintenance and repair, reliability is a key parameter. Table 1 summarizes some of the characteristics of the magnet power converters for LHC.

**Table 1:** Summary of LHC magnet power converters

| Power converter type | Quantity | Switch type[2] | ½ hour stability [ppm] |
|---|---|---|---|
| MB [13 kA, ±190 V] | 8 | SCR | 3 |
| MQ [13 kA, 18 V] | 16 | SM | 3 |
| Inner triplet [5–7 kA, 8 V] | 16 | SM | 5 |
| IPD and IPQ [4–6 kA, 8 V] | 174 | SM | 5 |
| 600 A type 1 [±0.6 kA/±10 V] | 400 | SM | 10 |
| 600 A type 2 [±0.6 kA/±40 V] | 37 | SM | 10 |
| 120 A [±120 A/±10 V] | 290 | SM | 50 |
| Orbit correctors [±60 A/±8 V] | 752 | SM | 50 |
| Warm magnets [1 kA/450–950 V] | 16 | SCR | 20 |

---

[2] SCR – Thyristor bridge; SM – Switched Mode

### 3.2 Ion sources

Ion sources (IS) facilities requires a low particle energy (hundreds of MeV). The dimensions are in the order of hundreds of metres. The accelerators can be of different types (cyclotrons or LINACs), also in particular configurations. Since both superconducting and conventional magnets are used, there is a large variety and number of power converters.

#### 3.2.1 FRIB (USA, under construction)

Citing the official web page of FRIB [13], it "will provide intense beams of rare isotopes—shortlived nuclei no longer found on Earth. FRIB will enable scientists to make discoveries about the properties of these rare isotopes in order to better understand the physics of nuclei, nuclear astrophysics, fundamental interactions, and applications for society. As the next-generation accelerator for conducting rare isotope experiments, FRIB will allow scientists to advance their search for answers to fundamental questions about nuclear structure, the origin of the elements in the cosmos, and the forces that shaped the evolution of the universe."

The layout is shown in Fig. 8. The facility uses linear accelerators organized in a 'paperclip' (or 'folded LINAC') structure to save space.

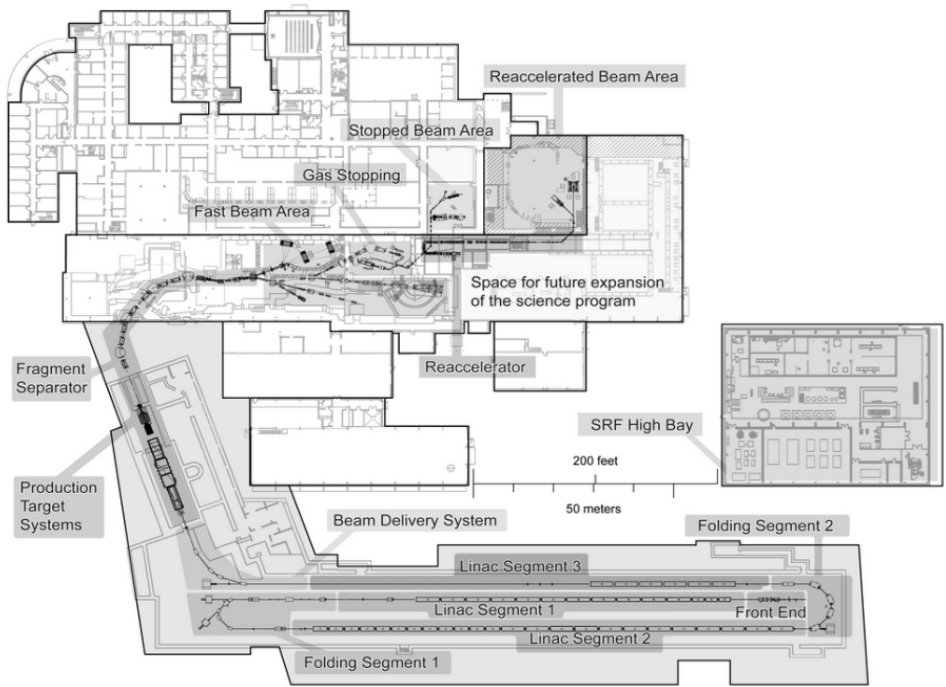

**Fig. 8**: FRIB layout (Credit: FRIB-MSU)

There is a large variety of magnetic—normal and superconducting—and electrostatic elements (dipoles and quadrupoles) to act on the particles, requiring either high voltages or high currents. From the power converter point of view, this translates in a mix of one quadrant (1-Q), two quadrant (2-Q) and four quadrant (4-Q) power converters (see Tables 2 and 3).

**Table 2:** FRIB electrostatic element power converters

| Quantity | $I_{out}$ [mA] | $V_{out}$ [kV] | Long-term stability [ppm] | Ripple [ppm] | Accuracy [ppm] | Resolution [ppm] |
|---|---|---|---|---|---|---|
| 11 | 1–60 | 1–100 | ±100–±500 | 100–200 | 1000 | 500 |

**Table 3:** FRIB magnetic element power converters

| Quantity | $I_{out}$ [A] | $V_{out}$ [V] | Long-term stability [ppm] | Ripple [ppm] | Accuracy [ppm] | Resolution [ppm] |
|---|---|---|---|---|---|---|
| 194 | 2–3500 | 6–600 | ±200–±1000 | 50–400 | 2500–4000 | 20–200 |

The relatively 'relaxed' parameters of some power converters allow the adoption of commercial units (so-called commercial-off-the-shelf (COTS), a solution that has some advantages in terms of costs, availability, reliability (units are produced in hundreds, over the years), maintenance and spares.

## 3.3 Cancer therapy

The centres for cancer therapy (CT) are, at the same time, an accelerator facility and a clinical facility: quite different environments, where technical, scientific, medical and psychological aspects coexist and have to be considered at the same time. This type of facility requires a low particle energy (hundreds of MeV). The dimensions are relatively small, in the order of tens of metres. The accelerators can be of different types (cyclotrons or synchrotrons). The power converters are of different types (SCR, PWM, linear) and their main requirements are high reliability (e.g. avoiding failures during treatment of patients) and minimization of repair time.

### 3.3.1 PROSCAN (PSI, Switzerland, 2007)

PROSCAN is a cyclotron-based facility for proton therapy within the Paul Scherrer Institute (PSI) in Villigen (see, for example, Ref. [14]); the layout is shown in Fig. 9. The proton beamlines are used for eye radiotherapy and deep-seated tumours. The proton source is the 250 MeV COMET cyclotron; its installation is shown in Fig. 10.

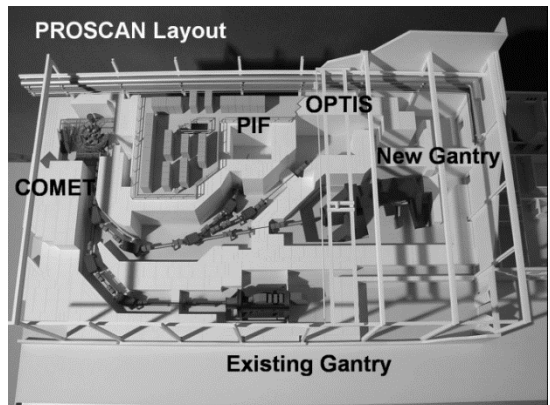

**Fig. 9**: PROSCAN Layout (Credit: PSI)

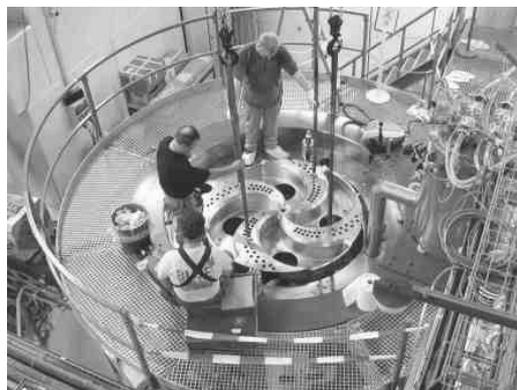

**Fig. 10**: The COMET cyclotron under construction in 2004 (Credit: PSI)

There are about 100 magnet power converters. They are all four-quadrant pulse-width modulation (PWM) type, and the main focus is on dynamics with tight requirements on d$i$/d$t$ and regulation delays. Table 4 summarizes the parameters of the PROSCAN magnet power converters.

**Table 4:** PROSCAN power converters

| Type | $I_{out}$ [A] | $V_{out}$ [V] | Long-term stability [ppm] | Ripple [ppm] | Reproducibility [ppm] | Accuracy [ppm] | d$i$/d$t$ [A/s] |
|---|---|---|---|---|---|---|---|
| IGBT | 500 | 350 | 100 | 50 | 100 | 500 | 125 |
| IGBT | 220 | 320 | | 1000 | | 1 000 | 11 000 |
| IGBT | 220 | 285 | 15 | 15 | 100 | 500 | 100 |
| IGBT | 150 | 175 | | 1000 | | 1 000 | 40 000 |
| IGBT | 150 | 90 | 100 | 50 | 100 | 500 | 100 |
| MOSFET | 50 | 50 | 500 | 100 | 500 | 500 | |
| MOSFET | 10 | 24 | 500 | 100 | 500 | 500 | |

### 3.3.2 CNAO (Italy, 2010 to 2012)

CNAO is a synchrotron accelerator (about 80 metres in circumference) where carbon ions or protons are accelerated and then driven to the treatment room. The protons are accelerated to 250 MeV while the carbon ions are accelerated to 480 MeV. Figure 11 shows the synchrotron ring.

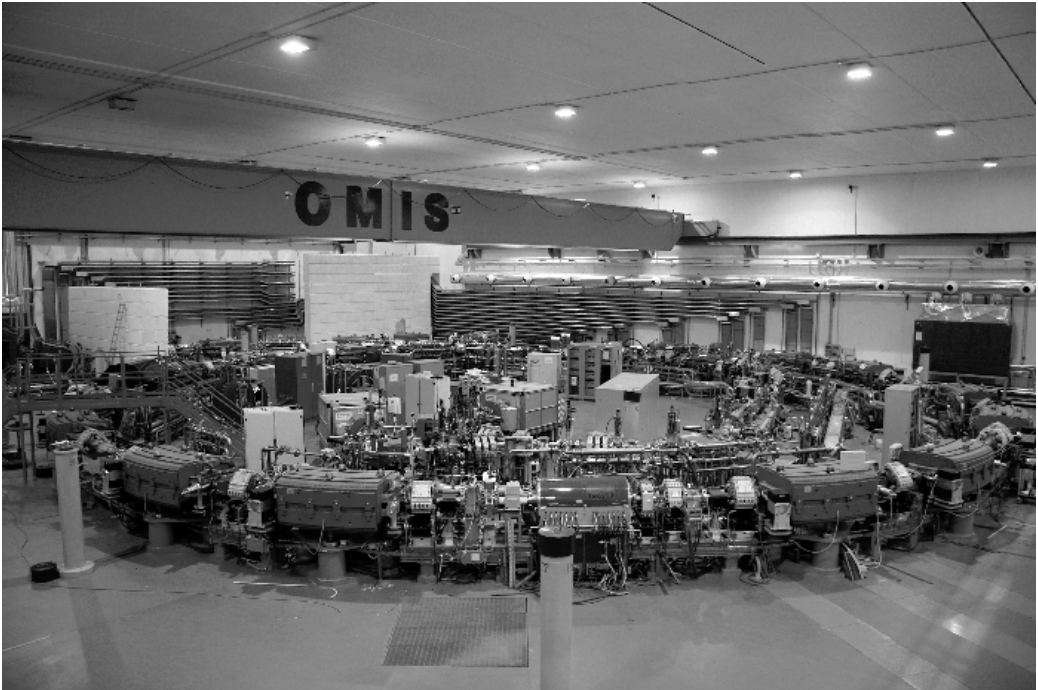

**Fig. 11**: CNAO synchrotron (Credit: CNAO)

Table 5 summarizes the characteristics of the more than 200 magnet power converters [15]. It has to be noted that there are high voltage and high current requirements at the same time—requiring the adoption of a mixed technology: thyristor (SCR) bridges with switched mode active filtering on the DC side of the bridges [16], and high reproducibility.

**Table 5:** CNAO power converters

| Magnet | Type | $I_{out}$ [A] | $V_{out}$ [V] | Long-term stability [ppm] | Ripple [ppm] | Reproducibility [ppm] | Resolution [ppm] |
|---|---|---|---|---|---|---|---|
| Dipole | SCR+SM AF | 3000 | ±1600 | ±5 | ±5 | ±2.5 | ±5 |
| Dipole | SCR+SM AF | 3000 | ±110 | ±25 | ±25 | ±13 | ±25 |
| Dipole | SCR+SM AF | 2500 | ±450 | ±5 | ±5 | ±2.5 | ±5 |
| Dipole | PWM | ±550 | ±660 | ±200 | ±100 | ±100 | ±60 |
| Dipole | PWM | ±30–300 | ±20–±35 | ±50–±500 | ±50–±250 | ±25–±500 | ±50–±1000 |
| Quadropole + sextupole | PWM | 150–650 | ±17–±65 | ±50–±100 | ±50–±100 | ±25–±50 | ±50–±100 |
| Corrector | Linear | ±30–±150 | ±15–±30 | ±500 | ±250 | ±500 | ±1000 |

### 3.3.3 *MedAustron (Austria, under construction)*

MedAustron uses a synchrotron accelerator (about 80 metres in circumference) with a LINAC pre-accelerator for the ions. Proton or carbon ions are accelerated and then driven to the treatment rooms. The protons are accelerated up to 250 MeV (800 MeV for non-clinical research) while the carbon ions are accelerated up to 400 MeV [17]. Figure 12 shows the facility building and its layout.

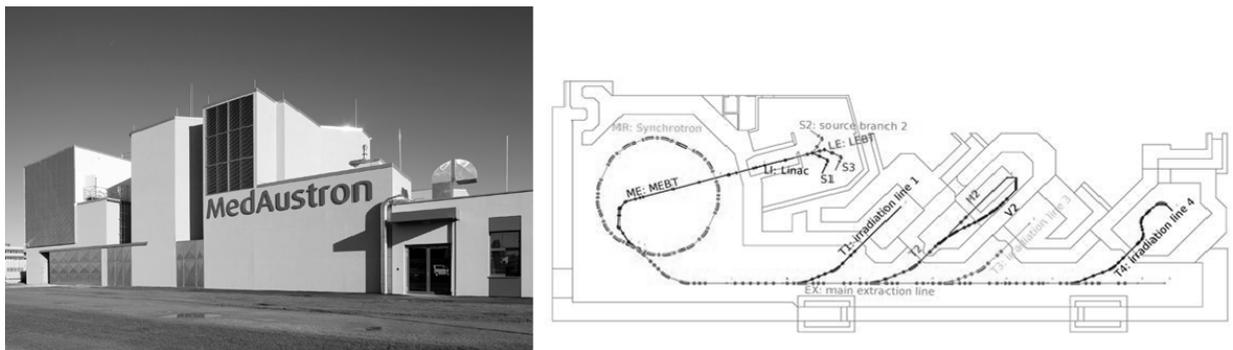

**Fig. 12**: MedAustron view and layout (Credit: MedAustron)

There are more than 200, four-quadrant, PWM power converters, operating at 0.5 Hz, with peak output power up to 4.5 MW. The precision range is between 10 ppm to 100 ppm. The power converters were specified to and provided by the manufacturers as voltage sources, while the high-precision current regulation system was designed and provided in collaboration by CERN and MedAustron [18].

### 3.4 Neutron sources

The production of neutrons from neutron sources (NS) requires linear or circular accelerators, of mid to low energy (0.8 GeV to 2.5 GeV) whose dimensions are in the order of some hundred metres or less. Neutrons are generated via spallation (a simple and clear description of the principle is given in Ref. [19]).

### 3.4.1 *ISIS (RAL, UK, 1985)*

ISIS [20] is a neutron source located in the Rutherford Appleton Laboratory (RAL) in the UK. It is based on a 800 MeV proton accelerator comprising a LINAC pre-accelerator and a synchrotron (165 m circumference) operated at 50 Hz, see Fig. 13.

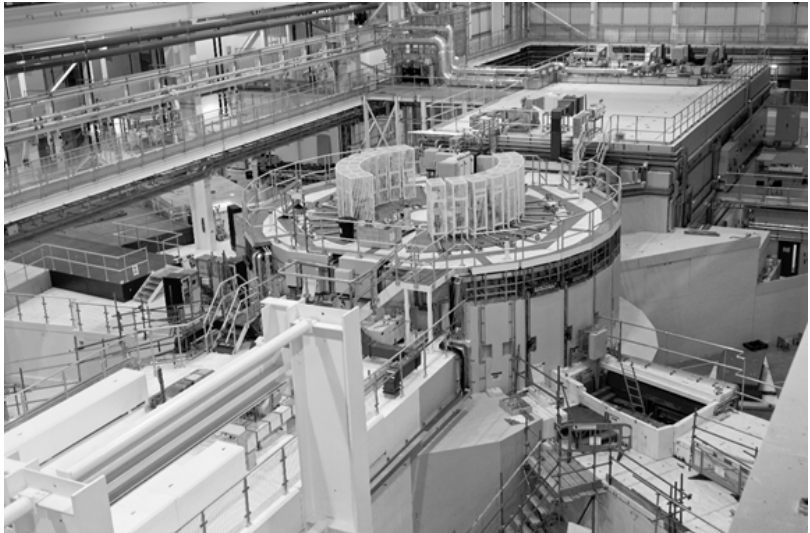

**Fig. 13**: ISIS (Credit: RAL, ISIS)

Operation at such a 'high' frequency, 50 Hz, imposes the adoption of a special configuration for the magnets and the associated power converters, the 'White circuit', from the name of the inventor (the original paper is Ref. [21]). At ISIS it consists of a 1 MJ resonant circuit with a DC bias power converter (660 A) and $4 \times 300$ kVA AC power converters, connected to a string of 10 chokes [22, 23], see Fig. 14. The rated secondary AC rms voltage is 14.4 kV at 1022 A.

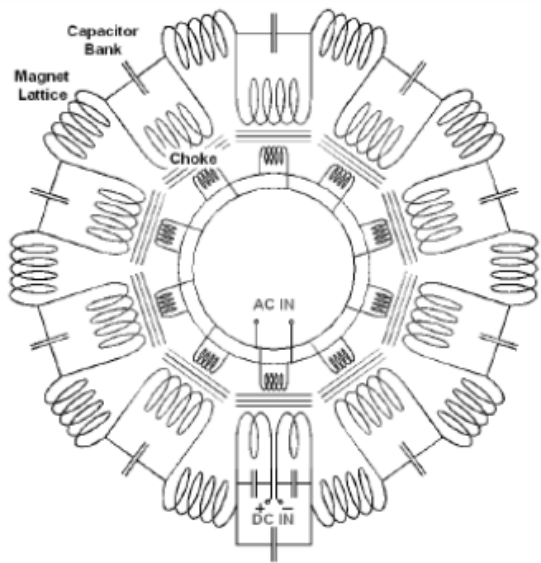

**Fig. 14**: ISIS White circuit layout from Ref. [22]

## 3.5 Light sources

Light sources (LS) utilize electrons to generate synchrotron radiation as a tool for research and are, usually, user-dedicated facilities (within an extremely vast literature, the properties of synchrotron radiation are described, in an unconventional but effective way, in Ref. [24]). They can be either 'circular'—the so-called 'storage rings'—or 'linear'—also known as the free electron laser (FEL).

Storage rings operate in the 1.5 GeV to 8 GeV range; they have dimensions of some hundreds of metres up to a few kilometres. They adopt either normal or superconducting magnets and—in particular for the most recent facilities—most of the magnets are individually energized, requiring a large number of power converters.

FELs operate at 1.5 GeV up to 20 GeV; the dimensions are ranging from some hundreds of metres up to a few kilometres; they use normal or superconducting magnets and they require a large number of power converters.

In the following paragraphs, I have briefly summarized the characteristics of the main magnet power converters of storage rings (SR, Sections 3.5.1 to 3.5.10) and free electron lasers (FEL, Sections 3.5.11 to 3.5.14). The year in parentheses by the name indicates the first stored electron beam or first electron beam along the whole structure.

### 3.5.1 Elettra (Italy, first e-beam 1993)

Elettra is a 'double energy facility': it regularly operates for users at 2.0 GeV (75% of the allocated beamtime) and 2.4 GeV (25% of the allocated beamtime) [25]. The storage ring circumference is 264 m and there are about 300 DC power converters adopting thyristor bridges for the main magnets and transistor-based bipolar linear topology for correctors [26, 27]. Table 6 summarizes the magnet power converters, while Fig. 15 is an aerial view of the storage ring building.

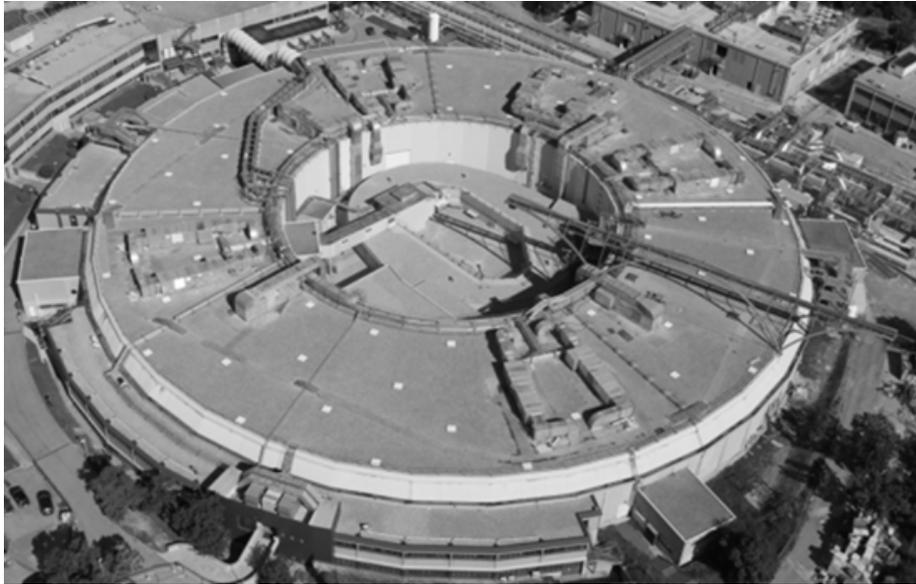

**Fig. 15**: Aerial view of the Elettra storage ring building (Credit: Elettra Sincrotrone Trieste)

**Table 6:** Elettra power converters

| Magnet | $I_{out}$ [A] | $V_{out}$ [V] | Long-term stability [ppm] | Ripple [ppm] | Resolution [ppm] | Type |
|---|---|---|---|---|---|---|
| Dipole SR | 2000 | 560 | ±200 | ±40 | 16 | SCR |
| Quadrupole SR | 300 | 560 | ±200 | ±40 | 16 | SCR |
| Sextupole SR | 300 | 560 | ±200 | ±40 | 16 | SCR |
| Corrector SR | ±16 | ±80 | ±500 | ±50 | 16 | Bipolar linear |

### 3.5.2 APS (USA, first e-beam 1995)

The Advanced Photon Source (APS) operates for users at 7 GeV [28]. The storage ring circumference is 1100 m and there are more than 1100 DC power converters adopting either thyristor bridges or PWM techniques. Table 7 summarizes the magnet power converters, while Fig. 16 is an aerial view of the facility.

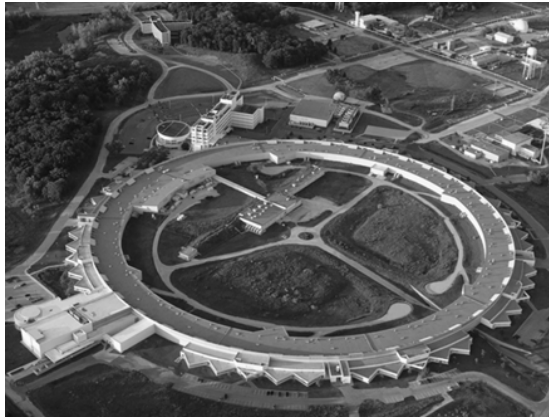

**Fig. 16**: Aerial view of APS (photograph Tigerhill Studio, Argonne National Laboratory)

**Table 7**: APS power converters

| Magnet | $I_{out}$ [A] | $V_{out}$ [V] | Long-term stability [ppm] | Ripple [ppm] | Resolution [ppm] |
|---|---|---|---|---|---|
| Dipole SR | 550 | 750 | ±30 | ±40 | 16 |
| Quadrupole SR | 500 | 20 | ±60 | ±800 | 16 |
| Sextupole SR | 250 | 25 | ±300 | ±200 | 16 |
| Corrector SR | ±150 | ±20 | ±30 | ±1000 | 16 |

### 3.5.3 LNLS (Brazil, first e-beam 1997)

The UVX machine at LNLS operates for users at 1.37 GeV. The storage ring circumference is 93 m and there are about 200 power converters (DC and AC for the booster) adopting either thyristor bridges or PWM techniques [29]. Table 8 summarizes the magnet power converters, while Fig. 17 is a view of the UVX.

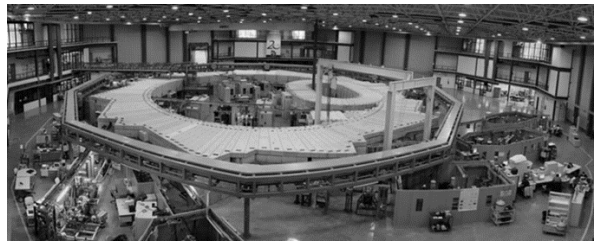

**Fig. 17**: View of UVX (Credit: LNLS)

**Table 8**: UVX (LNLS) power converters

| Magnet | $I_{out}$ [A] | $V_{out}$ [V] | Long-term stability [ppm] | Ripple [ppm] | Resolution [ppm] | Type |
|---|---|---|---|---|---|---|
| Dipole SR | 300 | 950 | ±100 | ±70 | 16 | SCR |
| Quadruple + sextupole SR | 10–220 | 10–45 | ±1000–±100 | ±1000–±100 | 16 | SCR |
| Corrector SR | ±10 | ±10 | ±1000 | ±100 | 16 | Bipolar linear |

### 3.5.4    SLS (PSI, Switzerland, first e-beam 2000)

The Swiss Light Source (SLS) operates for users at 2.4 GeV [30]. The storage ring circumference is 288 m and there are about 640 DC power converters. This is the first facility adopting an in-house developed fully digital regulation and control for the magnet power converters. Table 9 summarizes the magnet power converters, while Fig. 18 is a panoramic view inside the facility building.

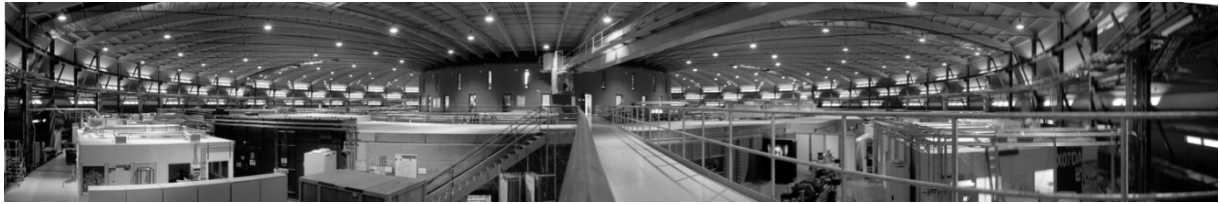

**Fig. 18**: View of SLS (Credit: PSI)

**Table 9:** SLS (PSI) power converters

| Magnet | $I_{out}$ [A] | $V_{out}$ [V] | Long-term stability [ppm] | Ripple [ppm] | Resolution [bit] | Accuracy [ppm] |
|---|---|---|---|---|---|---|
| Dipole SR | 500 | 880 | 100 | 15 | 16 | 100 |
| Quadruple + sextupole SR | 70–140 | 35–145 | 100 | 40–100 | 16 | 100 |
| Bipolar SR | ±7 | ±24 | 100 | 15 | 18 | 1000 |

### 3.5.5    SSRL-SPEAR3 (USA, first e-beam 2003)

SPEAR3 operates for users at 3.0 GeV [31]. The storage ring circumference is 234 m and there are about 250 DC power converters, including the transfer lines. The large power converters adopt both thyristor or diode bridges and PWM; while the small power converters are PWM [32, 33]. Table 10 summarizes the magnet power converters.

**Table 10:** SPEAR3 (SSRL) power converters

| Magnet | $I_{out}$ [A] | $V_{out}$ [V] | Long-term stability [ppm] | Type |
|---|---|---|---|---|
| Dipole SR | 800 | 1200 | 20 | 12p SCR bridge + PWM |
| Quadruple SR | 100 | 100–700 | 100 | 12p diode bridge + PWM/PWM |
| Sextupole SR | 225 | 600 | 100 | 12p diode bridge + PWM/PWM |
| Corrector SR | ±30 | ±50 | | PWM |
| Dipole TL | 500 | 45 | 100 | PWM |
| Quadruple TL | 60 | 80 | 100 | PWM |

### 3.5.6    Soleil (France, first e-beam 2006)

Soleil operates for users at 2.75 GeV. The storage ring circumference is 354 m and there are about 350 DC power converters, including the transfer lines. The adopted technologies are 12-pulse bridge with PWM output stage for dipole and sextupole power converters and PWM for quadrupoles and correctors [34]. Table 11 summarizes the magnet power converters, while Fig. 19 is a panoramic view of the facility site.

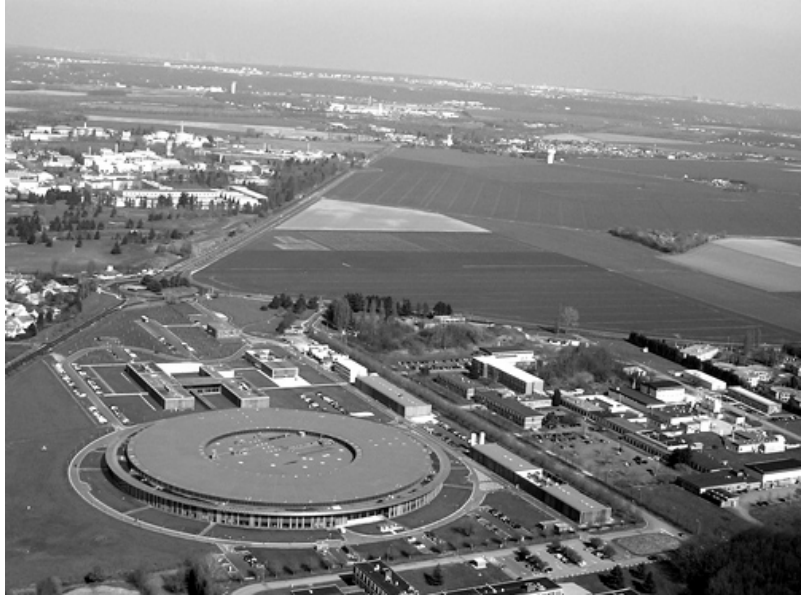

**Fig. 19**: View of Soleil (Credit: Soleil)

**Table 11:** Soleil power converters

| Magnet | $I_{out}$ [A] | $V_{out}$ [V] | Long-term stability [ppm] | Resolution [ppm] |
|---|---|---|---|---|
| Dipole SR | 580 | 610 | 10 | 10 |
| Quadruple + sextupole SR | 250–350 | 14–140 | 20–50 | 5–50 |
| Corrector SR | ±7–±14 | ±3.5–±14 | 20–50 | 2–30 |
| Dipole TL | 250–580 | 20–80 | 50–100 | 60–100 |
| Quadrupole TL | 10–275 | 9–10 | 50–100 | 20–60 |
| Corrector TL | ±1.5–±10 | ±2.5–±9 | 100–500 | 60–100 |

### 3.5.7  *Diamond (UK, first e-beam 2006)*

Diamond Light Source (DLS) operates for users at 3.0 GeV. The storage ring circumference is 560 m and there are about 1000 DC power converters [35]. The adopted technology is PWM with digital regulation. Table 12 summarizes the magnet power converters, while Fig. 20 is a panoramic view of the facility site.

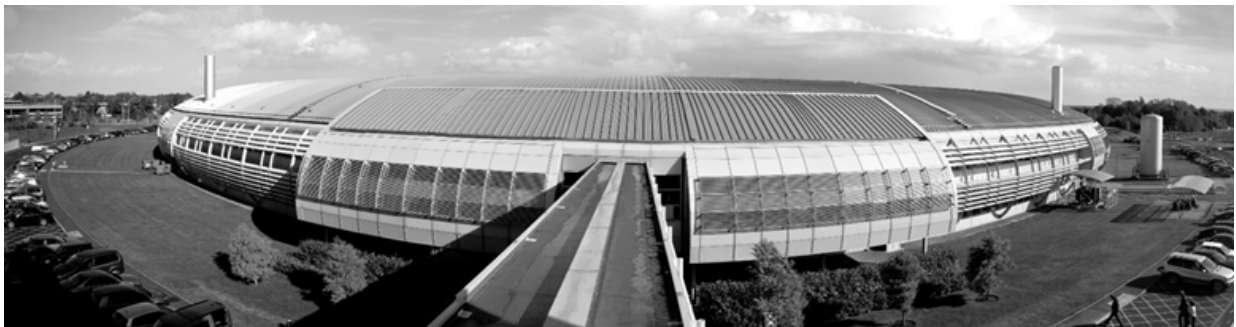

**Fig. 20**: View of Diamond (Credit: DLS)

**Table 12:** Diamond power converters

| Magnet | $I_{out}$ [A] | $V_{out}$ [V] | Long-term stability [ppm] | Ripple [ppm] | Resolution [ppm] | Reproducibility [ppm] | Bandwidth [Hz] |
|---|---|---|---|---|---|---|---|
| Dipole SR | 1500 | 530 | ±10 | ±10 | 4 | 10 | DC |
| Quadruple + sextupole SR | 100–350 | 17–41 | ±10 | ±10 | 4 | 10 | DC |
| Corrector SR | ±5 | ±20 | | | 4 | 10 | 50 |

### 3.5.8 ALBA (Spain, first e-beam 2010)

ALBA operates for users at 3.0 GeV. The storage ring circumference is 267 m and there are about 400 power converters (including those for the booster-based injector) [36]. The adopted technology is PWM with digital regulation. Table 13 summarizes the magnet power converters, while Fig. 21 is an aerial view of the facility site.

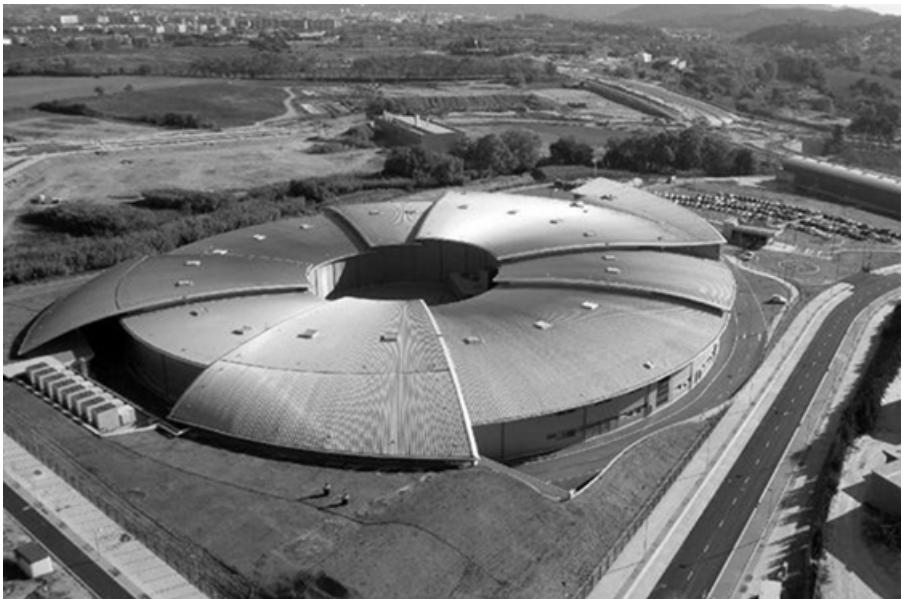

**Fig. 21**: View of ALBA (Credit: ALBA)

**Table 13:** ALBA power converters

| Magnet | $I_{out}$ [A] | $V_{out}$ [V] | Long-term stability [ppm] | Ripple [ppm] | Resolution [ppm] |
|---|---|---|---|---|---|
| Dipole SR | 600 | 750 | ±10 | 10 | 5 |
| Quadrupole SR | 200–225 | 15–25 | ±10 | 10 | 5 |
| Sextupole SR | 215 | 100–350 | ±50 | 50 | 15 |
| Corrector SR | ±12 | ±60 | ±20 | 10 | 5 |
| Dipole TL | 12–180 | 12–60 | ±15 | 15 | 15 |
| Quadrupole TL | 15–170 | 15–20 | ±15 | 15 | 15 |
| Corrector TL | ±2–±6 | ±2–±10 | ±100 | | 100 |

### 3.5.9    PETRA III (DESY, Germany, first e-beam 2010)

PETRA started as an electron-positron collider during the 1980s, then it was used as the pre-accelerator for the high-energy HERA ring. After the closing of the latter, PETRA has been converted into a light source, with the construction of experimental halls for hosting photon beam lines [37]. With its circumference of 2.3 km, PETRA III is one of the largest storage rings in the world. The magnet power converters adopt different technologies and topologies (there are White circuits for the ramped magnets) with a mix of analog and digital regulation.

Figure 22 shows an aerial view of the facility and Table 14 reports the characteristics of the main magnet power supplies.

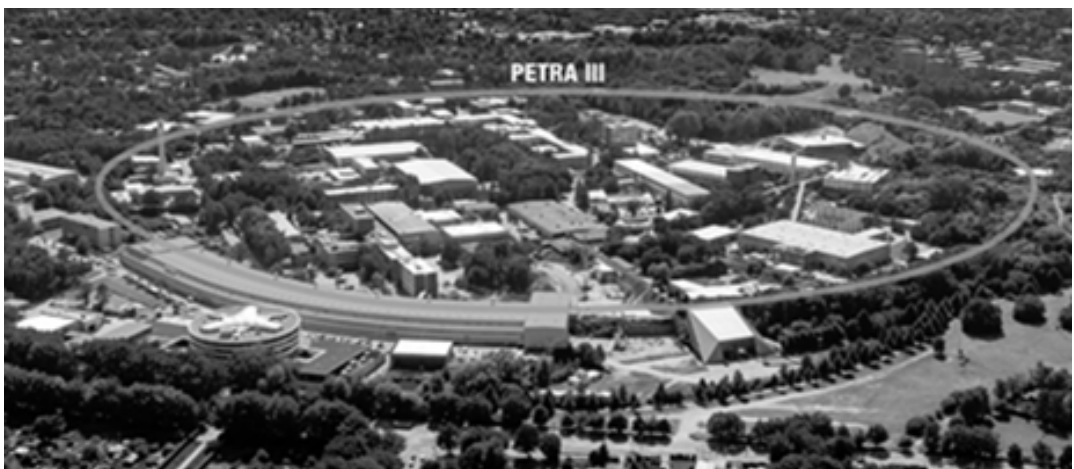

**Fig. 22**: PETRA III aerial view (Credit: DESY)

**Table 14:** PETRA III power converters

| Magnet | $I_{out}$ [A] | $V_{out}$ [V] | Ripple [$V_{out}$ rms] | Accuracy [ppm] | Resolution [bit] | Type |
|---|---|---|---|---|---|---|
| TL to PETRA | 200–400 | 60–120 | 2–3 | 100 | 16 | |
| Dipole AC | 1004 | 1330 | | 10 | 20 | White circuit |
| Dipole DC | 520 | 1560 | | 10 | 20 | White circuit |
| Quadrupole | 650 | 250 | | 10 | 20 | White circuit |
| Sextupole | 200 | 85 | | 10 | 20 | White circuit |
| Dipole PETRA | 600 | 300 | 3 | 100 (30) | 18 | SCR |
| Quadruple + sextupole PETRA | 200–600 | 60–120 | 2–3 | 100 (10) | 20 | PWM |
| Corrector PETRA | ±5–±55 | ±40–±60 | 2–3 | 500 (10) | 20 | PWM |

### 3.5.10    MaxIV (Sweden, under construction in 2014)

At the time of writing, MaxIV is under construction [38]. It is the most advanced circular light source, adopting innovative solutions, in particular for the magnets. The main storage ring (528 m in circumference) will operate at 3.0 GeV. A full-energy LINAC is the injector; and an intermediate, smaller storage ring, operating at 1.5 GeV, will be also part of the facility. Overall, about 1000 magnet power converters will be installed.

Figure 23 shows the construction of the main ring building while Table 15 reports the characteristics of the magnet power supplies for the MaxIV 3.0 GeV main ring.

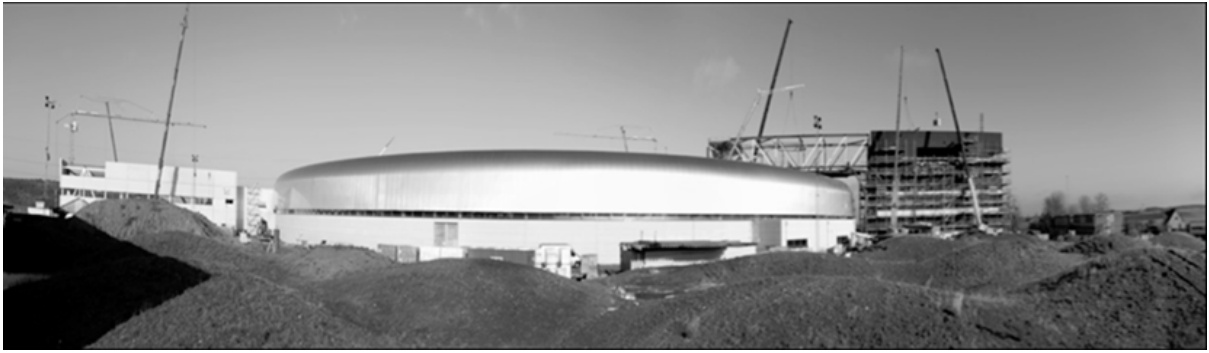

**Fig. 23**: MaxIV under construction (Credit: MaxLab)

**Table 15:** MaxIV main ring power converters

| Magnet (3 GeV) | $I_{out}$ [A] | $V_{out}$ [V] | Long-term stability [ppm] | Ripple [ppm] | Resolution [ppm] | Accuracy [ppm] |
|---|---|---|---|---|---|---|
| Main dipole | 450–750 | 145–210 | ±10 | ±10 | 16 | ±100 |
| Dipole strip | 175 | 44–80 | ±1000 | ±1000 | 16 | ±1000 |
| Quadrupole | 44–85 | 9–44 | ±100 | ±10.0 | 16 | ±1000 |
| Sextupole | 66–86 | 8–20 | ±100 | ±100 | 16 | ±1000 |
| Octupole | 58–217 | 45–104 | ±100 | ±100 | 16 | ±1000 |
| Corrector SR | ±5 | ±8 | ±25 | ±25 | 18 | |

### 3.5.11    LCLS (SLAC, USA, first e-beam 2009)

The final 1000 m of the 2.7 km-long SLAC LINAC are the accelerator for the Linac Coherent Light Source (LCLS), currently (2014), "the world's most powerful X-ray laser" [39], providing electrons at 13.6 GeV [40].

The magnet power converters parameters for the injector are reported in Table 16 [41]. Figure 24 shows some of the FEL undulators.

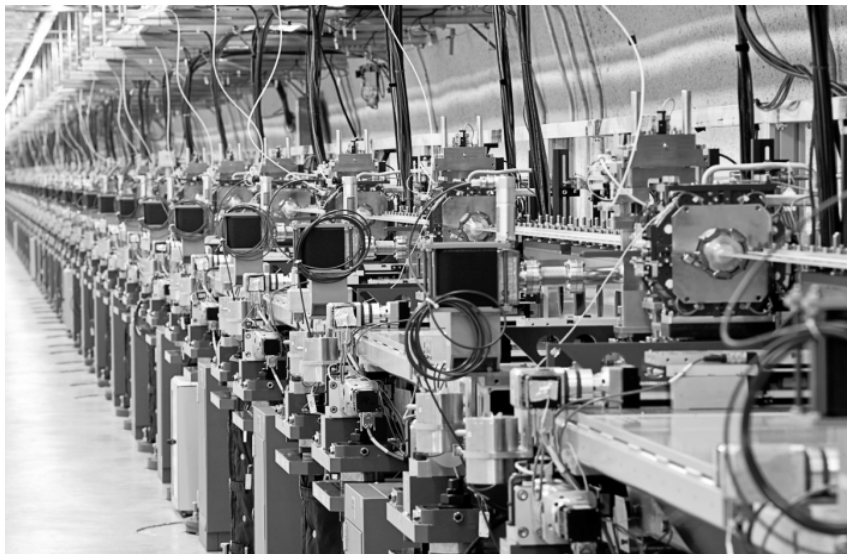

**Fig. 24**: Undulators at LCLS (Credit: LCLS-SLAC)

**Table 16:** LCLS power converters

| Magnet | $I_{out}$ [A] | $V_{out}$ [V] | Long-term stability [ppm] | Ripple [ppm] |
|---|---|---|---|---|
| Intermediate PC | Up to 375 | Up to 200 | 100 rms | 100 rms |
| Corrector | ±6–±30 | ±40 | 400 rms | 30 rms |

### 3.5.12   *FERMI (Italy, first e-beam 2011)*

After the construction and commissioning of the booster-based full injector for the Elettra storage ring in 2008, the 1 GeV LINAC and its plants have been re-used as part of the FERMI FEL source (360 m overall length). The construction of FERMI was completed in 2010, without affecting Elettra operation for external users, notwithstanding the proximity of the two sources, as can be seen in Fig. 25 (FERMI is the line stretching from the bottom to the top of the image on the left-hand side, almost tangential to Elettra's circular building). The energy of the electrons is 1.5 GeV and there are two undulators lines: FEL-1 to generate light in the vacuum UV (VUV) region (down to 20 nm), on the right-hand side in Fig. 27; FEL-2 provide light in the X-ray region between 20 to 4 nm (the undulators are on the left-hand side in Fig. 26) [42].

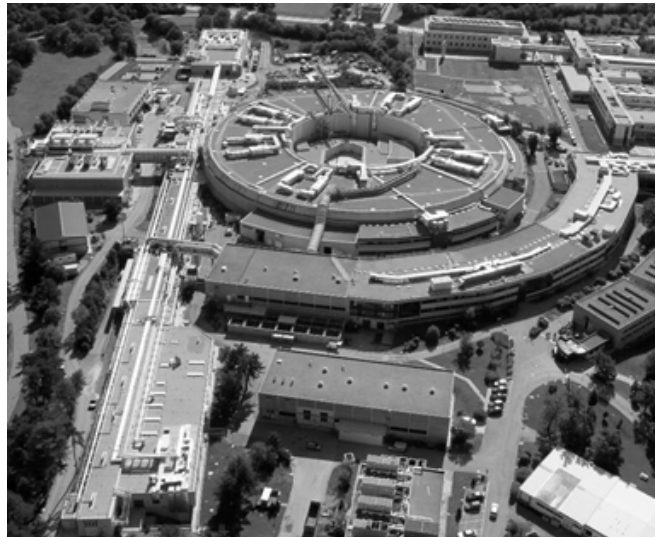

**Fig. 25**: Aerial view of FERMI and Elettra (Credit: Elettra Sincrotrone Trieste)

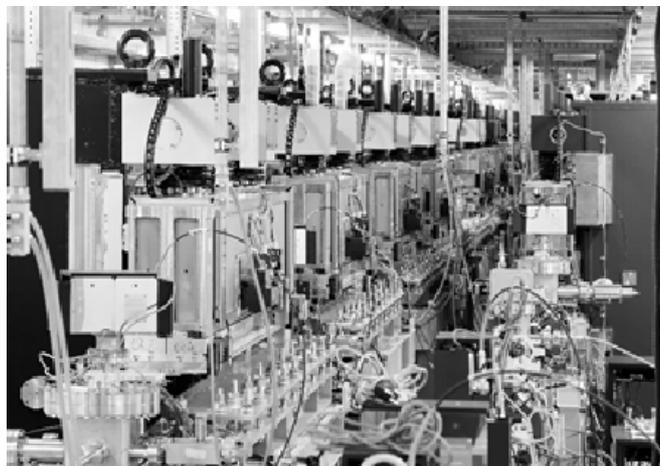

**Fig. 26**: FERMI undulators for FEL-1 (right) and FEL-2 (left) (Credit: Elettra Sincrotrone Trieste)

There are 37 different types of magnets and coils, supplied by 17 different types of DC power converters (~400 in total). A close collaboration between the magnet and the power converter teams could optimize the design, allowing the use of two types (in-house design) of power converters for 88% of the magnets. All power converters are of the PWM type and are custom-made and of commercial manufacture (COTS) and based on an internal design (bipolar ±5 A and ±20 A). Table 17 reports the characteristics for the magnet power converters [8, 43].

**Table 17:** FERMI power converters

| Magnet | $I_{out}$ [A] | $V_{out}$ [V] | Long-term stability [ppm] | Ripple [ppm] | Resolution [bit] |
|---|---|---|---|---|---|
| Dipole | 50–750 | 15–55 | 500 | 100 | 16 |
| Quadrupole | 5–100 | 10–30 | 25–500 | 10–100 | 16 |
| Corrector | ±5–±20 | ±10–±20 | 25–30 | 10–15 | 16 |

### 3.5.13    *SwissFEL (PSI, Switzerland, under construction in 2014)*

SwissFEL is the new source under construction (2014) at the Paul Scherrer Institut [44]. It is a free electron laser operating with electrons with a maximum energy of 5.8 GeV. The overall length is 740 m and there are ~600 magnet power converters. A feedback system involving correctors for the compensation of slow drifts is planned. Table 18 presents the main characteristics for the power converters.

**Table 18:** SwissFEL power converters

| Type | $I_{out}$ [A] | $V_{out}$ [V] | Ripple (10 Hz–30 kHz) [ppm] |
|---|---|---|---|
| 1-Quadrant IGBT | 220 | 40–100 | 50 |
| 4-Quadrant IGBT | ±150–±200 | ±40–±1100 | 3.5–50 |
| 4-Quadrant MOSFET | ±10–±50 | ±10–±20 | 10–100 |

### 3.5.14    *European XFEL (DESY, Germany, under construction in 2014)*

The European XFEL will stretch underground for about 3.4 km, accelerating electrons at 17.5 GeV (up to 20 GeV) with a superconducting LINAC [45]. The power converter will adopt different technologies and topologies, with digital regulation (analog for correctors).

Figure 27 shows the position of the European XFEL while Table 19 summarizes the parameters of the magnet power converter.

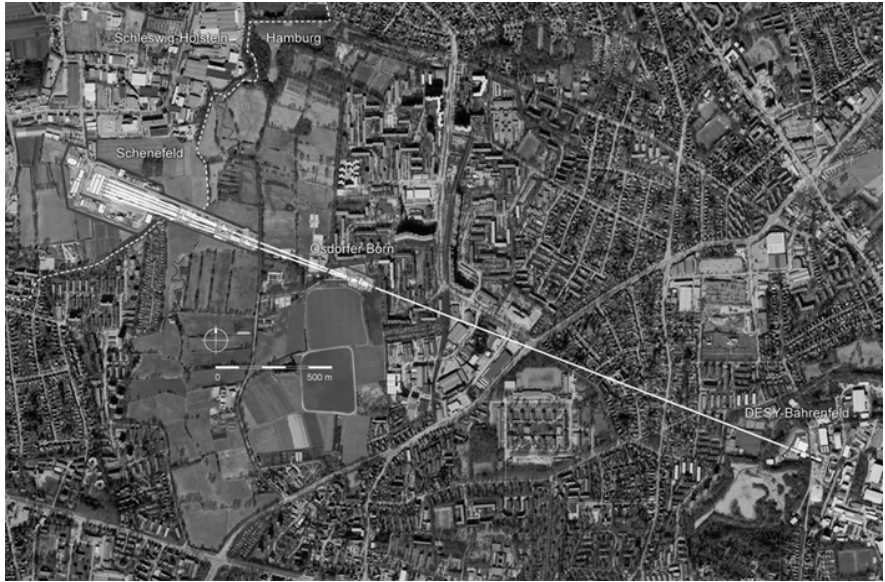

**Fig. 27**: XFEL position (Credit: European XFEL)

**Table 19:** European XFEL power converters

| Magnet | $I_{out}$ [A] | $V_{out}$ [V] | Ripple [$V_{out}$ rms] | Accuracy [ppm] | Resolution [bit] | Type |
|---|---|---|---|---|---|---|
| Main | 600–800 | 200–350 | 1% fs | 100 | 20 | SCR |
| Chopper | 200–600 | 60–120 | 1.5–3 | 100 | 20 | PWM |
| Small main | 5–10 | 40–60 | 2–3 | 100 | 20 | PWM |
| Corrector | ±5–±10 | ±40–±60 | 2–3 | 100 | 20 | PWM |

### 3.6 Booster synchrotrons and light sources

In modern 'circular' light sources, a gun, usually followed by a LINAC, emits and pre-accelerates the electrons, and then a 'booster synchrotron' increases their energy. There are exceptions, such as MaxIV [38], but normally a relatively small ring (compared to the main or storage ring) is used to ramp up the electrons to working energy. The booster usually operate at frequencies of 1–12 Hz. The current in the magnets—always positive, never changing sign—follows a sinusoidal waveform with a DC offset. Due to the inductance of the load and the cycling frequency, the power converters have to operate in two-quadrant mode for the output voltage, up to several kV.

From 1956 to 1998, the basic magnet/power converter structure for generating similar current/voltage waveforms has been the White circuit [21]. In 1998, at the EP2 Conference at ESRF, Grenoble, the PSI power converter team presented their design for the dipole power converter for the Swiss Light Source [46]. This solution adopted a digitally controlled and regulated, 1 MW peak, PWM power converter operating at 3 Hz, at a comparably 'low voltage', i.e. below 1 kV, providing the required current waveform to the magnets directly, without any resonating circuit. The cycling frequency is 'low'—below 5 Hz—but one of the main advantages is the possibility of generating non-sinusoidal waveforms for better matching the particle optics requirements [47] (the White circuit can only provide sinusoidal waveforms).

In the sections below I will provide a few examples of the power converters for booster magnets, before and after 1998, starting with the last installed White circuit in a European light source, BESSY II at the Helmholtz-Zentrum Berlin (HZB).

### 3.6.1 BESSY II (HZB, Germany, first e-beam 1998)

The structure of a White circuit is shown in Fig. 28, while the parameters for the White circuit power converters are summarized in Table 20 [48].

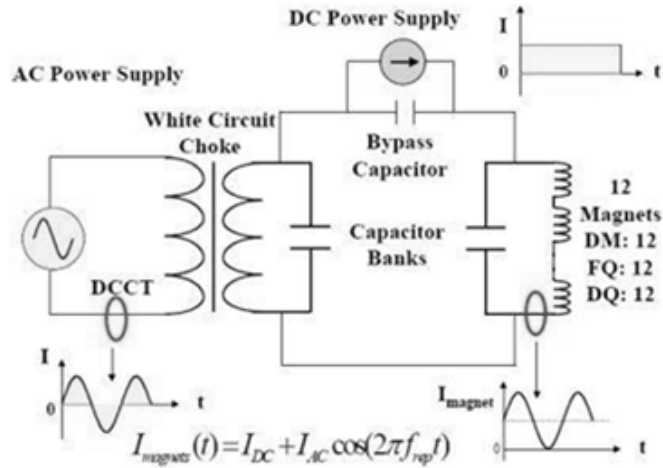

**Fig. 28**: White circuit schematics

**Table 20:** BESSY II White circuit power converters

| Magnet | $I_{out}$ [A] | $V_{out}$ [V] | Long-term stability [ppm] | Peak values on magnet circuit at 10 Hz |
|---|---|---|---|---|
| Dipole AC | 778 | 311 | | |
| Dipole DC | 1375 | 120 | ±20 | 2277 A / 3112 V |
| Quadrupole AC | 200 | 184 | | |
| Quadrupole DC | 340 | 70 | ±20 | 492 A / 527 V |

### 3.6.2 SLS (PSI, Switzerland, first e-beam 2000)

The SLS booster is the first implementation of a direct connection of power converters to the magnets. This is also the first example of digital control and regulation for the power converters. The working frequency is 3 Hz, the energy ramping is from 100 MeV to 2.7 GeV and there is one power converter for the dipoles [46]. Table 21 summarizes the main characteristics for the power converters.

**Table 21:** SLS booster power converters

| Magnet | $I_{out}$ [A] | $V_{out}$ [V] | Long-term stability [ppm] | Ripple [ppm] |
|---|---|---|---|---|
| Dipole | 950 | ±1000 | 100 | 10 |
| Quadrupole | 140 | ±120 | 100 | 100 |

### 3.6.3 LNLS (Brazil, first e-beam 2001)

The booster for UVX can be used also as a storage ring, at an intermediate energy. The repetition rate is 1 pulse every 6 s, ramping the energy of the electrons from 120 MeV up to 500 MeV. Table 22 reports the main parameters for the LNLS booster magnet power converter [49].

**Table 22:** LNLS booster power converters

| Magnet | $I_{out}$ [A] | $V_{out}$ [V] | Short-term stability [ppm] | Ripple [mA] |
|---|---|---|---|---|
| Dipole | 300 | 420 | 10 | ±120 |
| Quadrupole | 10 | 21 | 10 | ±10 |
| Sextupole | 10 | 26 | 10 | ±10 |
| Corrector | ±5–±6 | ±10 | 10 | ±1 |

### 3.6.4    Soleil (France, first e-beam 2001)

The Soleil booster has a repetition rate of 3 Hz, ramping the energy of the electrons from 100 MeV to 2.75 GeV. Due to the inductance of the magnets, in order to keep the peak voltage of the power converters below 1 kV, allowing the adoption of low-voltage techniques and rules, two separated power converters energize the dipole, with an 'twisted' connection of the coils, as reported in Fig. 29. Table 23 reports the main parameters for the magnet power converter of the Soleil booster [34].

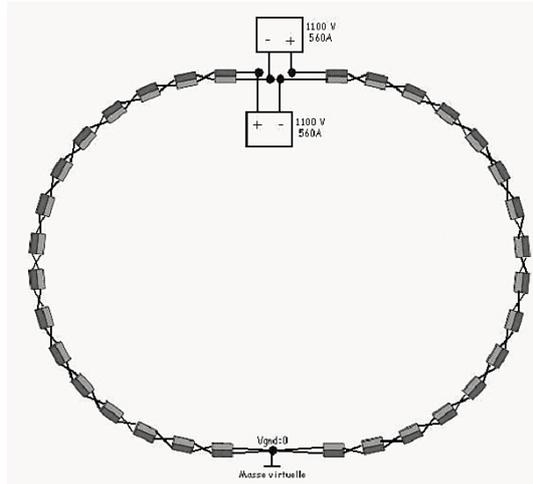

**Fig. 29**: Dipole magnet arrangement in the Soleil booster

**Table 23:** Soleil booster power converters

| Magnet | $I_{out}$ [A] | $V_{out}$ [V] | Accuracy [ppm] | Number of PCs |
|---|---|---|---|---|
| Dipole | ±580 | ±1000 | 50 | 2 |
| Quadrupole | ±250 | ±450 | 50 | 2 |
| Sextupole | ±30 | ±30 | 50 | 2 |
| Corrector | ±1.5 | ±2.5 | 50 | 44 |

### 3.6.5    Diamond (UK, first e-beam 2005)

The Diamond booster has a repetition rate of 3 Hz (5 Hz maximum), ramping the energy of the electrons from 100 MeV to 3 GeV. In this case, the dipole string was not split and there is only one power converter. Due to the inductance of the dipoles, the peak voltage is 2 kV, forcing the adoption of mid-voltage techniques and rules. The design of the power converters is modular, with redundancy, allowing continuity of operation even in the case of a single module failure.

Table 24 reports the main parameters for the Diamond booster magnet power converters [50].

**Table 24:** Diamond booster power converters

| Magnet | $I_{out}$ [A] | $V_{out}$ [V] | Reproducibility [ppm] | Resolution [ppm] | Number of PCs |
|---|---|---|---|---|---|
| Dipole | 1000 | ±2000 | ±50 | ±4 | 1 |
| Quadrupole | 200 | ±421 | ±50 | ±4 | 2 |
| Sextupole | 20 | ±60 | | | 2 |

### 3.6.6    Elettra (Italy, first e-beam 2007 to 2008)

A 1.0 GeV LINAC was the original injector for the Elettra storage ring. Between 2005 and 2007 a new full-energy injector was constructed and commissioned, without affecting the operations of the light sources for users [51]. Since March 2008 the 2.5 GeV booster, fed by a 100 MeV LINAC, is the full-energy injector for Elettra. Since 2010 Elettra operates in the so-called 'top-up' mode, keeping the current in the storage ring constant [25].

The Elettra booster has a circumference of 114 m, operates at 3 Hz, and, in a similar fashion to Soleil, there are two dipole magnet power converters, feeding the coils of the magnets separately, as shown in Fig. 30. An example of the actual current waveforms is given in Fig. 31. Table 25 summarizes the power converter parameters [52].

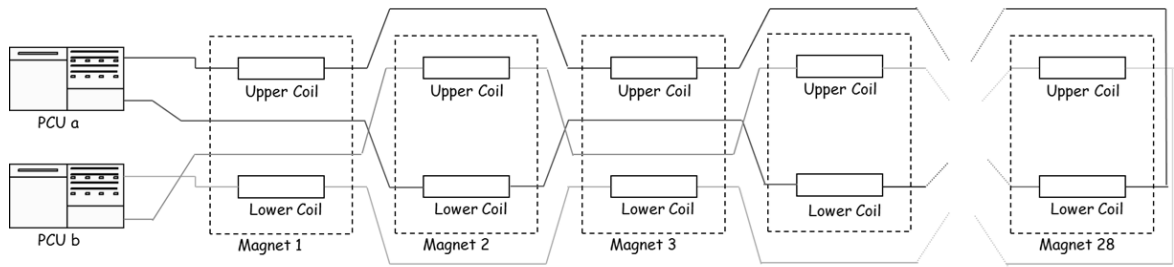

**Fig. 30**: Dipole magnet arrangement for the Elettra booster

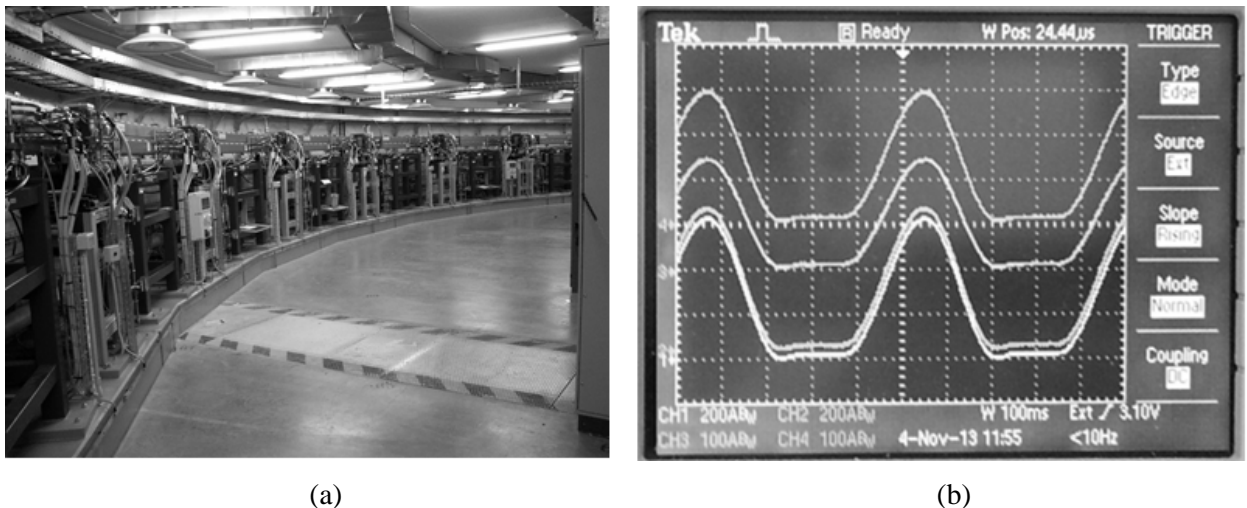

(a)                                                                                          (b)

**Fig. 31**: View of (a) the Elettra booster; (b) output current waveforms in the magnets. In (b) from top to bottom: quadrupole QD (100 A/division), quadrupole QF (100 A/division), dipoles 1 and 2 (200 A/division).

**Table 25:** Elettra booster power converters

| Magnet | $I_{out}$ [A] | $V_{out}$ [V] | Stability [ppm] | Number of PCs |
|---|---|---|---|---|
| Dipole | ±800 | ±1000 | ±5 | 2 |
| Quadrupole | ±400 | ±400 | ±50 | 2 |
| Sextupole | ±20 | ±35 | ±50 | 2 |
| Corrector | ±20 | ±50 | ±50 | 22 |

### 3.6.7    ALBA (Spain, first e-beam 2010)

The booster of ALBA has a repetition rate of 3 Hz, ramping the energy of the electrons from 100 MeV to 3 GeV. Also in this case, in a fashion similar to Soleil and Elettra, two separated power converters energize the dipole, with a 'twisted' connection of the coils

Table 26 reports the main parameters for the magnet power converter of the ALBA booster [53].

**Table 26:** ALBA booster power converters

| Magnet | $I_{out}$ [A] | $V_{out}$ [V] | Stability [ppm] | Resolution [ppm] | Reproducibility [ppm] | Number of PCs |
|---|---|---|---|---|---|---|
| Dipole | ±750 | ±1000 | ±15 | 5 | ±50 | 2 |
| Quadrupole | ±180 | ±120–±750 | ±15 | 5 | ±50 | 4 |
| Sextupole | ±8 | ±70 | ±50 | 15 | ±100 | 2 |
| Corrector | ±6 | ±12 | ±50 | 15 | ±100 | 72 |

## 4    Remarks and conclusions

In this paper, I have collected some examples of magnet power converters from quite a vast world of particle accelerators. It is quite straightforward that there are different requirements among particle accelerators for different applications. Maximum output current, its stability, its reproducibility, its accuracy and its dynamics ($di/dt$) are some of the most variable parameters. There are also significant differences among particle accelerators that have the same applications, according to the accelerator's type and age.

New technologies, both in the power converter field (components, low-level/local control) and in the feedback and remote control systems at a higher level, are also sources of change. In particular, new technologies allow extremely effective actions on the particle beam, like beam-based alignment (BBA) techniques for both circular ('multi-turn') machine and linear ('single-pass') ones[3]. The implementation of feedback systems—either on the 'orbit' or on the 'trajectory'—also benefits from new developments involving particle beam position monitors (BPM), very fast electronics, high computational power and a combination corrector magnet/power converter for compensating drift or oscillation of the particle beam with a bandwidth as large as possible.

---

3 The bibliography on BBA is quite vast: a simple search in the JACoW database [11] returned 64 results for titles containing the words 'beam-based alignment'

### 4.1 Storage ring correctors for light sources

As an example, in Table 27 I have collected the relevant data for the power converters for corrector magnets of the storage rings of eight synchrotron light sources. Two critical aspects of the corrector power converter assure smooth behaviour near 0 A: the so-called 'zero crossing' (0 A is a normal working point) and bandwidth. Bipolar, linear solutions can fulfil these requirements; the main drawback is poor efficiency. Such power supplies are still in use, at least at Elettra, and have been integrated into local and global 'fast feedback' system for stabilizing the particle beam (see, for example, Refs. [54, 55]).

A clear step was evident after 2000 with the adoption of PWM techniques for correctors, introducing digital regulation and a higher resolution. The output current is smaller, compared to the previous cases, because of an optimized design of the system magnet/power converter for the purpose and the characteristics of the accelerator (e.g. smaller transverse beam size, allowing a smaller gap in the magnet).

**Table 27:** Storage ring corrector power converters

| Facility (first year of operation) | Beam energy [GeV] | $I_{out}$ [A] | $V_{out}$ [V] | Long-term stability [ppm] | Ripple [ppm] | Resolution [bit] | Type |
|---|---|---|---|---|---|---|---|
| Elettra (1993) | 2.4 | ±16 | ±80 | ±500 | ±50 | 16 | Bipolar linear |
| APS (1995) | 7 | ±150 | ±20 | ±30 | ±1000 | 13 | |
| LNLS (1997) | 1.37 | ±10 | ±10 | ±1000 | ±100 | 16 | Bipolar linear |
| SLS (2000) | 2.4 | ±7 | ±24 | 100 | 15 | 18 | PWM |
| Soleil (2006) | 2.75 | ±7–±14 | ±3.5–±14 | 20–50 | | 16–18 | PWM |
| DLS (2006) | 3 | ±5 | ±20 | | | 18 | PWM |
| ALBA (2010) | 3 | ±12 | ±60 | ±20 | 10 | 18 | PWM |
| Max IV (…) | 3 | ±5 | ±8 | ±25 | ±25 | 18 | PWM |

### Acknowledgements

My gratitude goes to several colleagues, both at Elettra and from facilities worldwide for their support and information provided while collecting the material for my work: M. Cautero and T. Ciesla (Elettra), H.-J. Eckoldt (DESY), K. Holland (FRIB), R. Kuenzi (PSI), S. Murphy (ISIS), R. Petrocelli (ALBA), C. Rodriguez (LNLS), P. Tavares (MaxIV) and J. Wang (APS)

I also wish to thank the following companies for having provided me—anonymously, for reasons of confidentiality—material on the power supplies they have built for particle accelerators: Bruker/SigmaPhi Electronics, EEI and OCEM.

*Collaboration: BIW, COOL, CYCLOTRONS, DIPAC, ECRIS, FEL, HIAT, ICALEPCS, ICAP, ICFA ABDW, IPAC, LINAC, NA-PAC, PCaPAC, RuPAC and SRF".*

I collected most of articles cited in the following references from the JACoW database, through its search engine. It is sufficient to insert the author or the title to retrieve the electronic version of the article of interest.

**Bibliography**

D. Brandt, Ed., Specialized course on power converters, CERN 2006–010 (2006).

A.W. Chao, K.H. Mess, F. Zimmerman and M. Tigner, Eds, *Handbook of Accelerator Physics and Engineering*, (World Scientific Publishing Company, 2013), 2nd ed.

K.A. Olive et al. (Particle Data Group), Chin. Phys. C, **38**, 090001 (2014).
http://pdg.lbl.gov/2014/reviews/contents_sports.html

**Useful web links**

Joint Accelerator Conferences website: http://www.jacow.org/index.php?n=Main.Proceedings

Synchrotron light facilities: http://www.lightsources.org/regions

High Energy Physics facilities: http://www.interactions.org/cms/?pid=1000025

IEEEXplore Digital Library (available for IEEE members or purchase):
http://ieeexplore.ieee.org/xpl/conferences.jsp

European Power Electronics and Drives Conferences (available for download to EPE members or purchase): http://www.epe-association.org/epe/index.php

Test Infrastructure and Accelerator Research Area (TIARA): http://www.eu-tiara.eu

Accelerators for Society: http://www.accelerators-for-society.org